\documentclass[journal]{IEEEtran}
\usepackage{cite}

\ifCLASSINFOpdf
   \usepackage[pdftex]{graphicx}
\else
\fi

\usepackage{algorithm}
\usepackage{algpseudocode}  
\usepackage{comment}  
\usepackage{graphicx}
\usepackage{array}
\usepackage{color,xcolor}
\usepackage{setspace}
\usepackage{threeparttable}
\usepackage{amsfonts}
\usepackage{amsmath}
\usepackage{bm}
\usepackage{xcolor}
\usepackage{circledsteps}
\usepackage{makecell}

\usepackage{amssymb}
\usepackage{multirow}
\usepackage{pict2e}
\usepackage{tabularx}
\usepackage{newfloat}
\usepackage{listings}

\begin{document}

\title{MA\textsuperscript{2}: A Self-Supervised and Motion Augmenting Autoencoder for Gait-Based Automatic Disease Detection}

\author{\normalsize{
\thanks{Yiqun Liu is with School of Information Technology and Media, Hexi University, Zhangye, 734000, P.R. China, Email:liullhappy888327@163.com.} 
Yiqun Liu,
\thanks{Ke Zhang
is with Transportation Research Center,Tsinghua University, Email: zhangkethu@mail.tsinghua.edu.cn.}Ke Zhang,
\thanks{Yin Zhu is with Shanghai Key Laboratory of Intelligent Information Processing, School of Computer Science, Fudan University, Shanghai, 200433, P.R. China, Email: yinzhu20@fudan.edu.cn}Yin Zhu,
}

\thanks{This paper has supplementary downloadable source code available at https://github.com/EchoItLiu/MA2-PyTorch provided by the authors.}}

\maketitle

\begin{abstract}
Ground reaction force (GRF) is the force exerted by the ground on a body in contact with it. GRF-based automatic disease detection (ADD) has become an emerging medical diagnosis method, which aims to learn and identify disease patterns corresponding to different gait pressures based on deep learning methods. Although existing ADD methods can save doctors time in making diagnoses, training deep models still struggles with the cost caused by the labeling engineering for a large number of gait diagnostic data for subjects. On the other hand, the accuracy of the deep model under the unified benchmark GRF dataset and the generalization ability on scalable gait datasets need to be further improved. To address these issues, we propose MA\textsuperscript{2}, a GRF-based self-supervised and motion augmenting autoencoder, which innovatively models the ADD task as an encoder-decoder paradigm. In the encoder, we introduce an embedding block including the 3-layer 1D convolution for extracting the token and a mask generator to randomly mask out the sequence of tokens to maximize the model's potential to capture high-level, discriminative, intrinsic representations. whereafter, the decoder utilizes this information to reconstruct the pixel sequence of the origin input and calculate the reconstruction loss to optimize the network. Moreover, the backbone of an autoencoder is multi-head self-attention that can consider the global information of the token from the input, not just the local neighborhood. This allows the model to capture generalized contextual information. Extensive experiments demonstrate MA\textsuperscript{2} has SOTA performance of $90.91\%$ accuracy on $1\%$ limited pathological GRF samples with labels, and good generalization ability of $78.57\%$ accuracy on scalable Parkinson’s disease dataset. See supplementary materials for source code, and two pre-processed datasets are downloadable at https://github.com/EchoItLiu/MA2-PyTorch.
\end{abstract}

\section{Introduction} \label{Sec1}

\IEEEPARstart{B}{ioinformatics} is regarded as an important branch of modern biology and a powerful tool. In more detail, it takes advantage of biometrics to analyze various types of bioinformatics characteristics for revealing the mysteries of life activities. In recent years, in the field of bioinformatics for disease, indicators of bioinformatics have been playing an increasingly important role. That is to say, data-driven bioanalysis provides new standpoints and insights for the application of disease prevention, diagnostic precision, adjuvant therapy, disease surveillance, and personalized medical plans. Therefore, this interdisciplinary scientific field integrates the knowledge and technologies of multiple disciplines, including biometrics and computer science, as well as scholars in related fields who carry out the collection, storage, analysis, and interpretation of biometric data, which provides unprecedented opportunities for the medical field.

Gait is one of the most important and unique biological characteristics of the human body. It refers to the distinct behavior pattern displayed by people during walking, which involves the trait of force, direction, and is the instantiation of people's intrinsic walking habits during the stages of landing, starting, stance, and swing. Neurologists, physiotherapists, and orthopedic surgeons have all been working on the analysis of gait for a long time, so as to understand an individual's health status and potential neurological or muscle diseases. However, these methods mainly rely on medical experience and are therefore not objective enough. Recently, by virtue of the accelerated growth of intelligent computing power, AI-based computational methods of bioanalysis have ushered in a development by leaps and bounds, the use of AI technology to automate the analysis of polymorphic bioinformation data such as gait data, which is mainly used to improve the detection accuracy of diseases.

Gait data can be mainly divided into two categories: non-sensor data and sensor data. Among them, non-sensor data, including the video image and depth cameras, respectively. Besides, it has long-distance and non-cooperative characteristics during the collection process. On the other hand, sensor-based data, refers to the gait information directly collected by the devices installed in specific locations, such as the pelma, belts, joints, etc. These instruments can be accelerometers, gyroscopes, and plantar pressure sensors, which can be used for sampling the mechanical data generated during every stage of gait in real time. For example, ground reaction force (GRF), which is used as the key bioinformatic feature to describe and analyze gait. We can measure it based on our interaction with the ground. A person standing motionless on the ground exerts a contact force on it (equal to the person’s weight), and at the same time, an equal and opposite ground reaction force is exerted by the ground on the person. In summary, non-sensor or image-based gait technology is more suitable for identification and tracking, while gait feature information collected by sensors, especially GRF, has broader application prospects and potential in automatic disease detection (ADD), making it more suitable to build non-invasive tools for the classification of disease in clinical settings.

For ADD methods based on GRF, some machine learning (ML) algorithms have been proposed in recent years. Slijepcevic et al.~\cite{2022Explaining} investigate the helpful and explainable ML method with layer-wise relevance propagation (LRP), which can increase transparency in automated clinical gait classification based on a gait dataset comprising GRF measurements, and the experimental results are nearly consistent with those assessed by clinicians. Next, Teixeira et al.~\cite{teixeira2022automatic} put forward an extreme machine learning (ELM) for GRF gait disease detection to classify healthy and pathological gait on GaitRec~\cite{2020GaiTRec} datasets, a large-scale ground reaction force dataset. Compared with linear regression, EML has the advantages of few training parameters, fast inference speed, and strong generalization ability. Furthermore, Siddiqui et al.~\cite{siddiqui2024footwear} summarize four machine learning techniques as forecast primary methods, such as random forest for classifying the lower limb disorders affecting the knee, hip, and ankle joints from Tehsil Head Quarter hospital Sadiqabad. Meanwhile, the study emphasized the importance of preprocessing and feature extraction before analyzing the collected data. Since deep learning technologies~\cite{yuan2016feature} achieved outstanding results in the ImageNet Large Scale Visual Recognition Challenge (ILSVRC) with its end-to-end feature extraction capabilities, researchers have begun to consider how to employ these advanced technologies to GRF non-invasive disease detection, that is, extracting deeper, discriminative, and individual-differentiated features~\cite{horst2023modeling} from GRF data to improve the performance of prediction. Boompelli et al.~\cite{2021Design} focus on the development of a telemetric gait analysis insole that works in conjunction with a mobile application and convolutional neural network to differentiate between healed and healing patients from the Austrian Workers’ Compensation Board (AUVA) and GaitRec. However, the 1D-CNN used in this article lacks sufficient comparative experiments with machine learning methods to demonstrate its advancement. Therefore, GaiRec-Net~\cite{pandey2022gaitrec} was proposed for binary classification on the healthy control and gait disorders based on GaitRec and compared with three machine learning methods. The experiment proves that the proposed deep learning model is better for feature extraction, resulting in high accuracy. In addition to that, Yun~\cite{yun2023exploratory} explores the application of residual network (ResNet) framework to human gait analysis for classification of lower limb injury on GaitRec dataset augumented by TimeGAN~\cite{yoon2019time}. Experiments show that the resnet architecture outperformed unsupervised clustering models—self-organizing maps and k-means clustering by a large margin, despite the propensity for overfitting. 

Overall, deep learning networks for ADD in GRF have broad, developing prospects. However, to the best of our knowledge, there are still the following challenges to be addressed: 

\begin{enumerate}
\item \textbf{\textit{Uniform and standardized benchmark datasets still lack:}} Until now, GRF-based disease detection has utilized datasets from multiple hospitals and research institutes~\cite{eltoukhy2017prediction}, but they are not unified and proprecessed with different modes. Hence, this makes it difficult to fairly evaluate the performance of different approaches, especially the computational advantages demonstrated by deep learning methods.

\item \textbf{\textit{It's not aligned with real clinical applications:}} Pathological gait samples (positive samples) are usually very scarce and more difficult to collect than healthy samples (negative samples). On the other hand, pathological samples must be diagnosed and evaluated by experienced physicians before they can be labeled, so they come with a high cost and are often poorly labeled~\cite{al2023pad}. Clinically, the main problem we face is how to train efficient deep models from fewer, and inadequately labeled pathological gait samples. 

\item \textbf{\textit{The flexibility of the model is limited:}} Models trained with supervised GRF samples may not be able to adequately learn the ``intrinsic features" from the data, i.e., the implicit patterns and structures, and therefore are not suitable for the detection of various diseases~\cite{sreeraman2023drug}. To this end, we should consider exploring advanced methods based on self-supervised or unsupervised learning to improve the generalization of the model. 
\end{enumerate}

To solve the problems above, we propose a pre-trained ADD method called motion augmenting autoencoder (MA\textsuperscript{2}) and capture the high-level, discriminative locomotion of gait from limited pathological GRF-based samples with labels, thereby, enhancing the performance of classification. Concretely, our contributions to this paper are summarized into the following three parts: 

\begin{itemize}
\item \textbf{\textit{GRF benchmarks in the real world:}} We integrated two canonical GRF datasets, GaitRec~\cite{2020GaiTRec} and Gutenberg~\cite{horst2021gutenberg}, as new benchmark datasets. Besides, from the perspective of real-world clinical application scenarios, we adjusted the ratio in the pathologic and healthy samples and performed channel-level minimum-maximum normalization~\cite{jlassi2024effect}. Meanwhile, the multiple deep learning methods~\cite{2021Design, pandey2022gaitrec,yun2023exploratory} for ADD in GRF are listed and conduct the experiments, so that researchers can reproduce these deep models in the future and make fair contrasts with each other.
    
\item \textbf{\textit{The SOTA ADD method based on GRF:}} We proposed a self-supervised deep model called motion augmenting autoencoder (MA\textsuperscript{2}) for extracting gait-based motion features from GRF data. The model adopts a masked auto-encoder structure based on Vision Transformer (ViT) for enhancing its ability to representation, and then fine-tuning with a small number of labeled samples to obtain a well-trained encoder. Multiple qualitative and quantitative experiments prove that MA\textsuperscript{2} can achieve an accuracy of over 90.91$\%$ for ADD with only 1$\%$ of labeled samples, which is beyond the reach of other comparative methods.  

\item \textbf{\textit{The satisfactory generalization:}} We transfer all the comparison methods and the proposed MA\textsuperscript{2} to a brand new GRF-based gait Parkinson’s disease detection dataset from the National Institutes of Health (NIH), and the experiments show that MA\textsuperscript{2} still achieves the best detection accuracy 78.57$\%$ under the new dataset, indicating that MA\textsuperscript{2} has a better performance in terms of generalization.
\end{itemize}

\section{Related Work}\label{sec2}

{\bf GRF-based disease gait detection} \quad The Ground Reaction Force (GRF)-based deep learning algorithm for disease gait detection has been a hotspot in biomechanics and medical fields in recent years. GRF, as a force generated by the interaction between the human body and the ground, provides complete gait information about the subjects, which is of great significance for disease screening, evaluation, and rehabilitation. Furthermore, GRF gait features can reflect the force of the human body interacting with the ground during walking, running, and other modes of exercise, including vertical, anterior-posterior, and internal and external components, which can provide abundant gait motions to establish statistical models. Many studies have applied deep learning algorithms to disease gait detection based on GRF. These studies have successfully identified gait features associated with a variety of diseases such as Parkinson's disease~\cite{andrei2019parkinson,fuadah2022parkinson,alharthi2020gait}, musculoskeletal impairments~\cite{chakraborty2022musculoskeletal,harithasan2023review,wu2020imu,sivakumar2022instrumented}, and etc.

{\bf Deep autoencoder} \quad It is a self-supervised deep network designed to learn a compressed representation of input data (encoding), and reconstruct it into the original data (decoding) via a decoder. Until now, the classical deep autoencoders that have been proposed include i) sparse autoencoders~\cite{ng2011sparse} through encoding sparsity constraint is introduced to avoid overfitting;  ii) Cheng et al.~\cite{cheng2018deep} proposed the convolutional autoencoders that combine convolutional networks with autoencoders to process image data; iii) Doersch  et al. put forward a variational autoencoders~\cite{doersch2016tutorial} to generate new data samples by introducing randomness; in addition, iv) Vincent et al. proposed a denoising autoencoder~\cite{vincent2010stacked} to train the model by adding noise to the input data to enhance the ability of expression; v) Kaiming et al. proposed asked autoEncoders (MAE)~\cite{he2022masked} designed an asymmetric encoder-decoder structure. Among them, the encoder only processes the visible part of the input image (the unmasked area), whereas the decoder combines the output of the encoder and the mask information to reconstruct the full pixel of the original image . 

In this study, we propose a GRF-based autoencoder called MA\textsuperscript{2} by integrating the MAE and 1D convolutional embedding layers to capture the underlying representations from limited GRF samples with labels. Specifically, a two-step learning strategy based on pre-training and fine-tuning was adopted from MAE to improve the representation ability of the autoencoder itself and the generalization ability for the downstream task, respectively. MA\textsuperscript{2} follows the mask strategy of MAE; that is, the pretext task is set to mask the input tokens in a high proportion so that only visible tokens are encoded, which can stimulate the model to learn more rich and informative representations to the maximum extent, thus improving the model performance and at the same time increasing the training speed.

\section{Methodology} \label{ml_methods}

\begin{figure*}[!htb]
\centering
\includegraphics[width=0.825\textwidth]{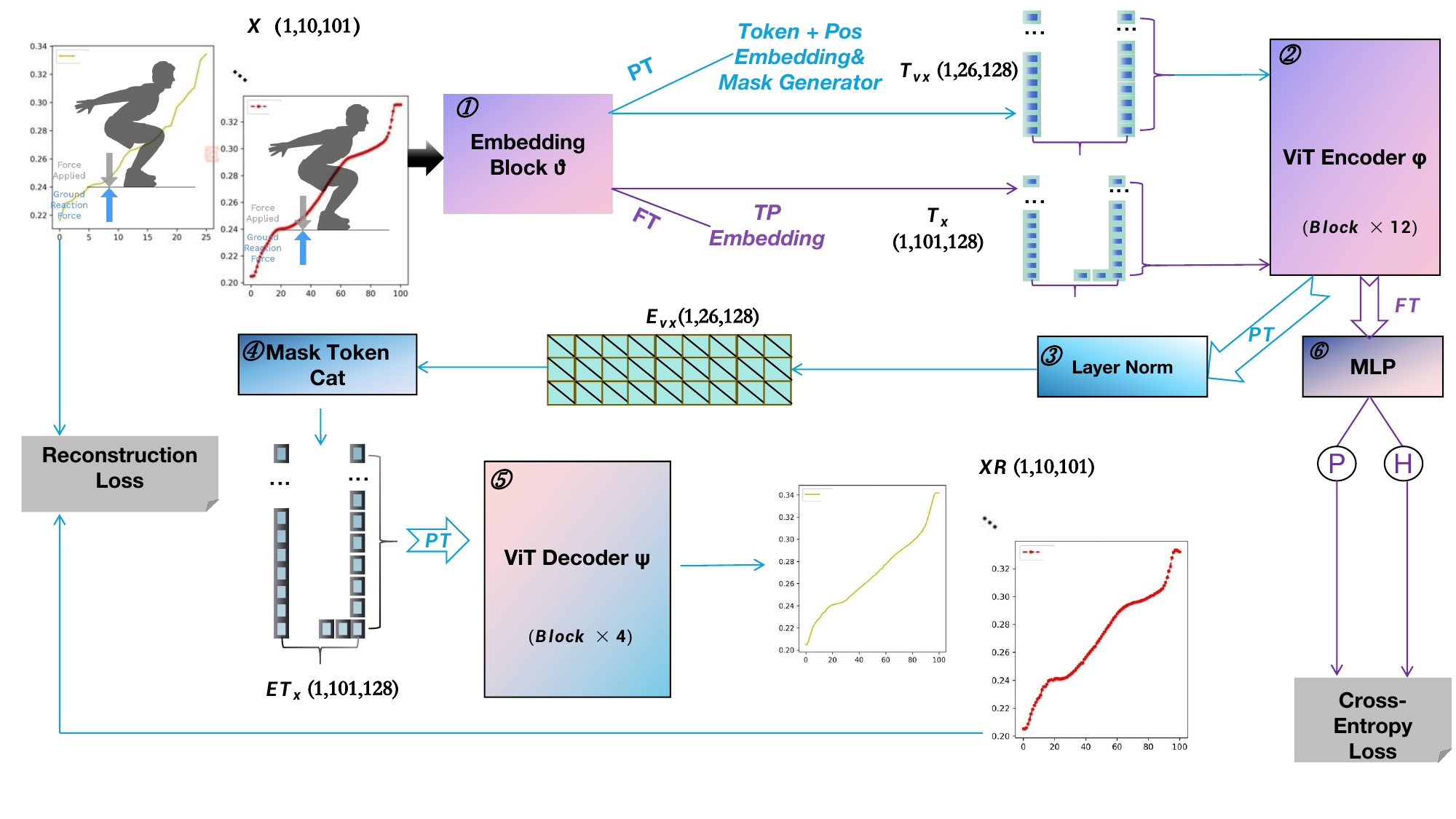}
\caption{{\bf The framework of the MA\textsuperscript{2}}. The framework mainly conducts automatic disease detection (ADD) based on GRF through two steps in sequence: i) pre-training (PT); ii) fine-tuning (FT). Concretely, the gait input data is $\mathbf{X}\in\mathbb{R}^{B \times C \times \ell}$, where $B$ is the batch size with an initial value of 1. $C$ (numbers of channels) is 10 and $\ell$ (sequence length) is 101. In the pre-training phase, $\mathbf{X}$ is utilized for capturing the high-level representations and then reconstructing the output $\mathbf{XR}$ with the same shape of input through module \Circled{1} Embedding Block $\rightarrow$ \Circled{2} ViT Encoder Block $\rightarrow$  \Circled{3} Layer Norm~\cite{ioffe2015batch} $\rightarrow$ \Circled{4} Mask Token $\rightarrow$ \Circled{5} ViT Decoder Block. On the other hand, in the fine-tuning phase, validation data $\mathbf{X}$ can only pass the \Circled{1} Embedding Block $\rightarrow$ \Circled{2} ViT Encoder Block $\rightarrow$ \Circled{6} MLP projector for inferring the health status ({\bf P}athological/{\bf H}ealthy) of the subject by plantar gait pressure. Figure \ref{pic3} and Figure \ref{pic4} show the operation details of module \Circled{1}. Similarly, the framework construction and operation details of Block \Circled{2} ViT Encoder Block and \Circled{4} Mask Token are shown in Figure \ref{pic56} and Figure \ref{pic7}, respectively.}
\label{pic1}
\end{figure*}
In this section, we will introduce the proposed motion augmenting autoencoder (MA\textsuperscript{2}) framework and discuss all modules of the two-phase training strategy of the framework in detail. The whole architecture of MA\textsuperscript{2} and pre-training/fine-tuning processes are shown in Figure \ref{pic1}.

\subsection{Preliminaries of MA\textsuperscript{2}}
MA\textsuperscript{2} is an enhanced autoencoder mainly comprising an 1D convolutional embedding block $\mathbf{\vartheta}$, a Vision Transformer (ViT)~\cite{alexey2020image} encoder $\mathbf{\phi}$, and a ViT decoder $\mathbf{\psi}$ that can be applied to extract the desired features for CV or NLP tasks.The detailed pipeline for pre-training and fine-tuning can be found in the supplementary material.

\subsection{Model Architecture}
\begin{figure}[!htb]
\centering
\includegraphics[width=0.9\columnwidth]{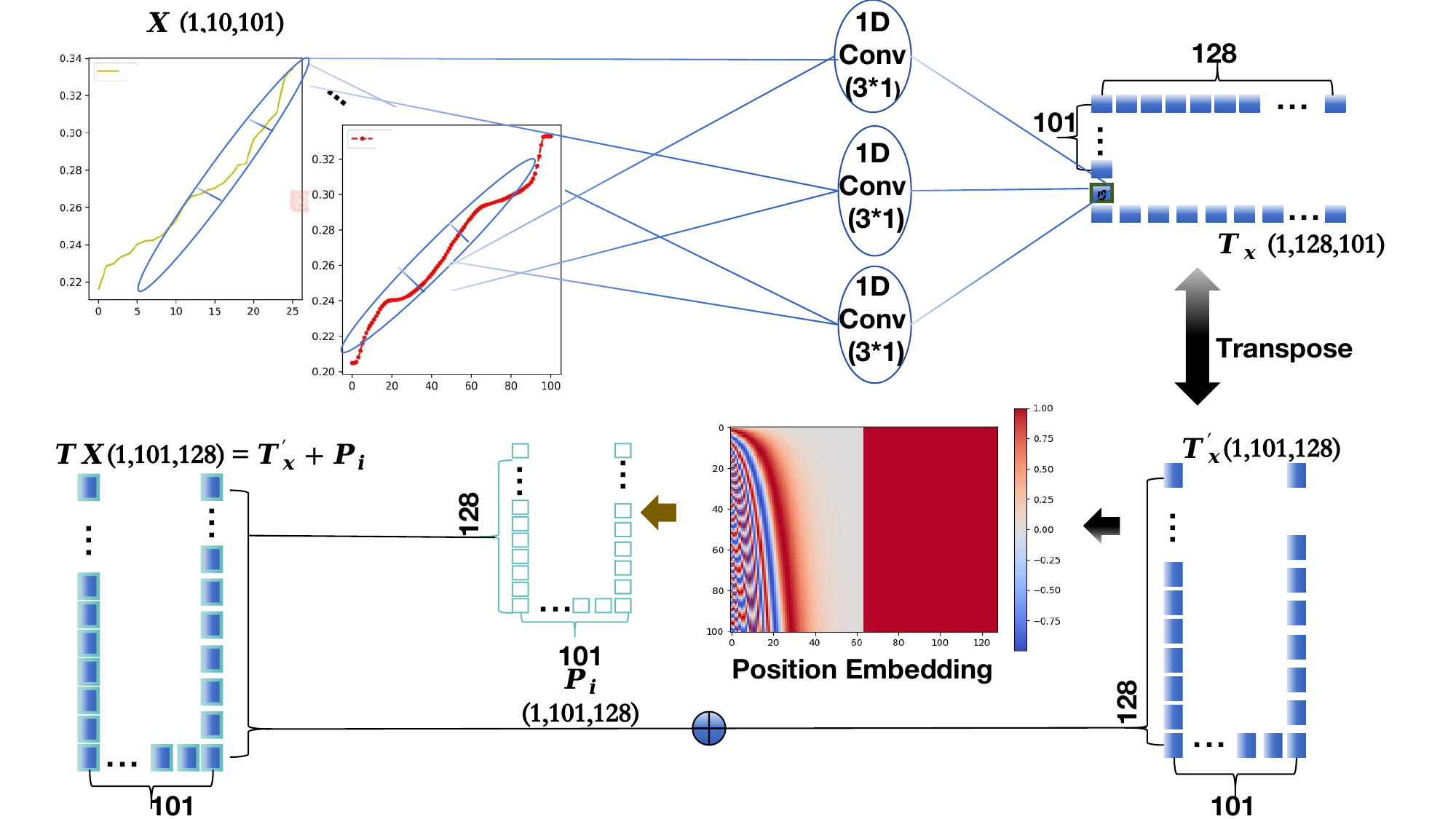}
\caption{\textbf{The process of token embedding and position embedding:} the original input data $\mathbf{X}$ is obtained by 1D convolution with a kernel size of 3 $\times$ 1 to obtain tokens $\mathbf{T}_{x}$ and transpose them to obtain tokens $\mathbf{T}_{x}'$, then the position embedding $\mathbf{P}_{i}$ is obtained by the sine-cosine encoding, and finally $\mathbf{T}_{x}'$ is added to obtain the tokens $\mathbf{TX}$.}
\label{pic3}
\end{figure}

Given an input data $\mathbf{X}$, i) the $\mathbf{\vartheta}$ mainly maps the $\mathbf{X}$ into a higher space as tokens $\mathbf{T}_{x}= \mathbf{\vartheta(X)}$. For the GRF-based gait dataset, $\mathbf{X}$ can be defined as a measurement of the plantar pressure of gait at multiple consecutive points, and $\mathbf{X}\in\mathbb{R}^{C \times \ell}$, where $C$ is the channel number and $\ell$ is the point number that is the length of the gait sequence. Besides, a mask generator in $\mathbf{\vartheta}$ is used to mask out a certain percentage of $\mathbf{T}_{x}\in\mathbb{R}^{\ell \times C_{1}}$, where $C_{1}$ is the number of channels of $\mathbf{T}_{x}$ ; ii) $\mathbf{\phi}$ encodes only visible points ($\mathbf{T}_{vx}$ that is not covered) and does not process mask tokens $\mathbf{T}_{mx}$, then generating potential representations of these visible tokens $\mathbf{E}_{v} =\mathbf{\phi}(\mathbf{T}_{vx})$ as outputs of the encoder; iii) as for $\mathbf{\psi}$, it is a lightweight transformer stack that concatenates the outputs of the encoder $\mathbf{E}_{v}$ and mask tokens $\mathbf{T}_{mx}$ as inputs and reconstructs the invisible parts previously masked by the generator of $\mathbf{\vartheta}$, finally outputs the reconstructed $\mathbf{XR}$. Specifically, the reconstruction loss is defined using pixel-wise mean square error (PMSE) loss as shown in Eq~\ref{rec_loss}
\begin{equation}\label{rec_loss}
\mathcal{L}_{PMSE} = (\frac{1}{n} * \Sigma(\mathbf{X} - \mathbf{XR})^2),
\end{equation}where $n$ is the number of pixels.

\begin{figure}[t]
\centering
\includegraphics[width=0.9\columnwidth]{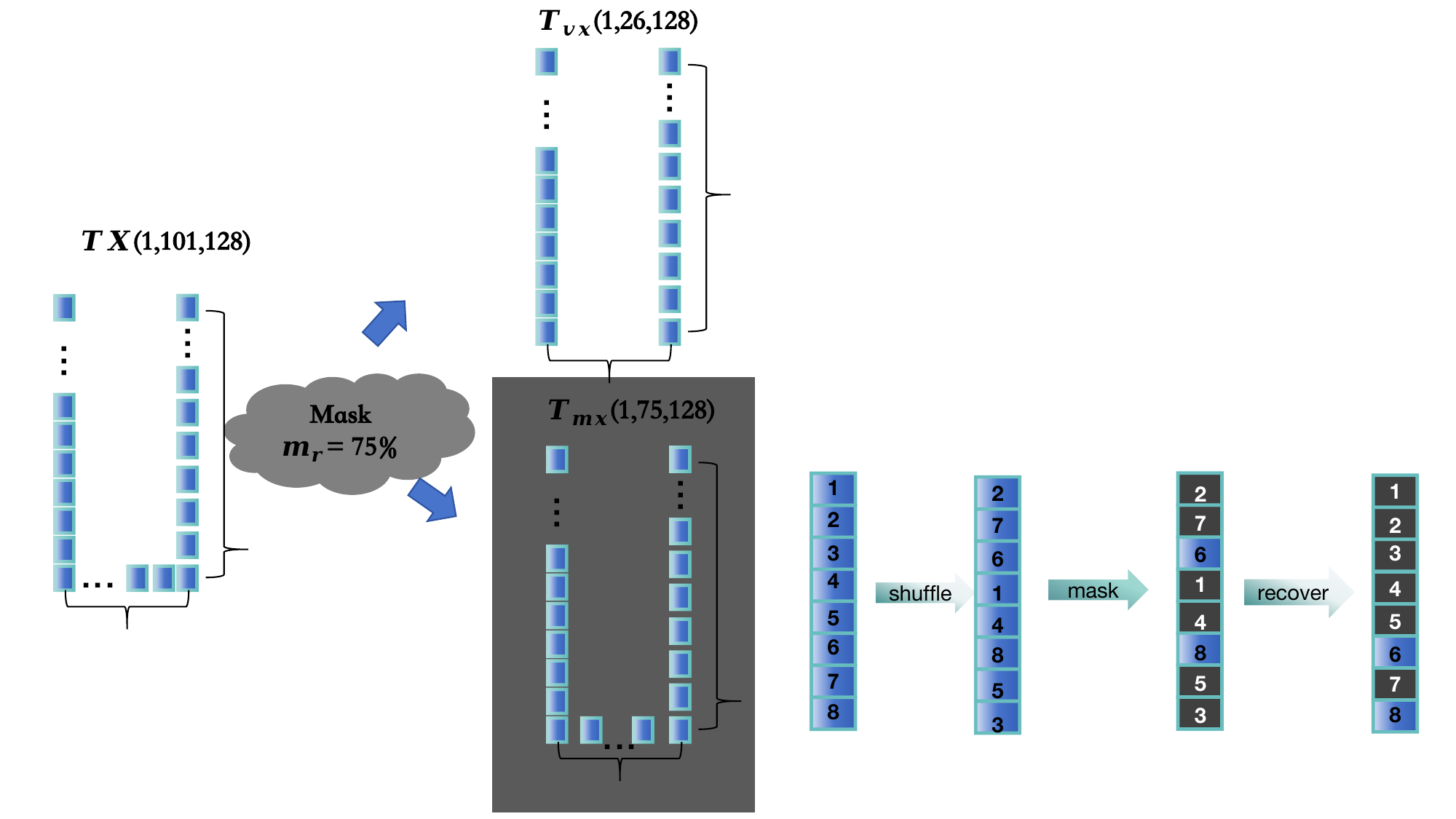}
\caption{\textbf{The mask process for the token is shown above:} First, set a fixed mask rate of $m_r=75\%$, and then randomly shuffle the 101 token sequences, take the first $101 \times (1-m_{r})=26$ feature vectors (only briefly expressed as 6 and 8 in the figure), and restore the original order of the 26 token vectors. These 26 vectors are fed into the model for the next encoder training, thus masking out the remaining 75 vectors.}
\label{pic4}
\end{figure}

{\bf Embedding Block} \quad The embedding block consists of multiple 1D convolutional layers, position encoding, and mask generators. Following the pipeline shown in Figure \ref{pic3}, 1D convolution is mainly used to extract shallow features from GRF-based data and acquire the tokens $\mathbf{T}_{x}$. Different from the patch embedding~\cite{alexey2020image} for the vision task, slicing and flattening of the image are required. For GRF gait data, input data $\mathbf{X}$ is equivalent to the intermediate result of executing the two steps mentioned above, and only transpose is needed. Second, we know that each pressure for gait at multiple consecutive points is as sequential as the sentence in the text. Therefore, we need to add the position information for each token. Here, we use sine and cosine functions with different frequencies to calculate the position information $\mathbf{P}i$:
\begin{eqnarray}\label{position}
\begin{split}
\mathbf{P}i_{(pos,2i)} & = sin(\frac{pos}{10000^{2i/d_{p}}})\;, \\
\mathbf{P}i_{(pos,2i+1)} & = cos(\frac{pos}{10000^{2i/d_{p}}})\;,\\
\mathbf{TX} & = \mathbf{T}_{x}' + \mathbf{P}i\;,
\end{split}
\end{eqnarray}where $pos$ is the position of the sequence for each token and $i$ is the sequential index. Here, $2i$ and $2i+1$ represent the even and odd positions, respectively. $d_{p}$ is a constant that represents the number of dimensions into which the position encoding is embedded in the space. The last inputs $\mathbf{TX}$ into the encoder should be the position information $\mathbf{P}i$ (visual position information) plus the tokens $\mathbf{T}_{x}'$. Third, the mask generator sets a mask rate to randomly mask out part of the token sequence $\mathbf{TX}$ and then obtains the visible sequence $\mathbf{T}_{vx}$ illustrated in Figure \ref{pic4}. The masked sequences $\mathbf{T}_{mx}$ will no longer be directly observed by the encoder, but position information of $\mathbf{T}_{mx}$ will be retained.

{\bf ViT Encoder} \quad $\mathbf{T}_{vx}$ are fed into a chain of blocks based on transformer for encoding. A detailed description of structure of the block can be found in the supplementary materials. In Figure \ref{pic56}, a stacked multi-head self-attention module forms a ViT encoder $\mathbf{\phi}$ to capture high-level and globally contextual representations $\mathbf{E}_{vx}$. That is, these features are calculated only from the unmasked $\mathbf{T}_{vx}$.

\begin{figure}[t] 
\centering
\includegraphics[width=0.9\columnwidth]{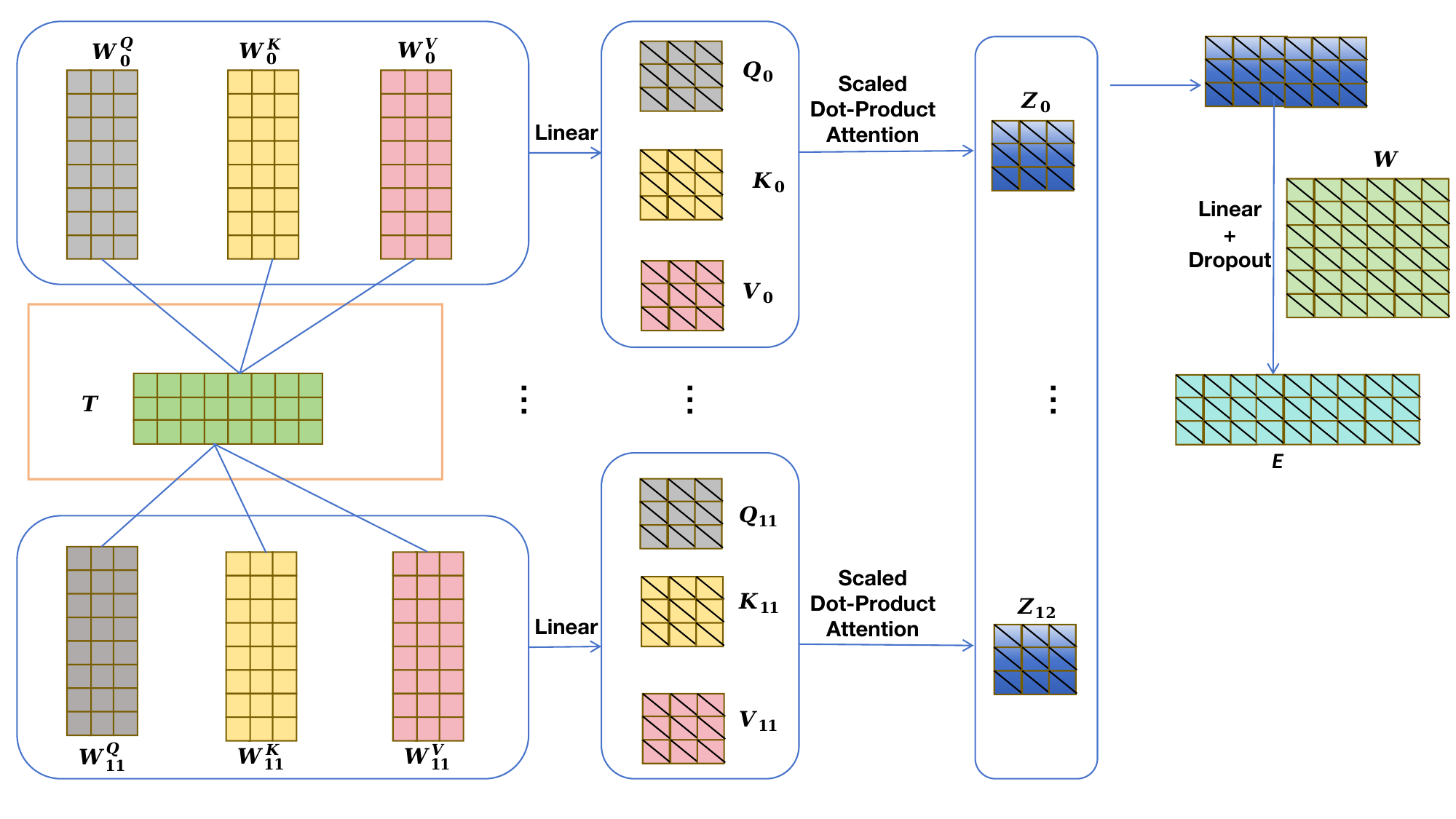}
\caption{multi-head attention mechanisms: Input $\mathbf{T}_{vx}$ in $h$ mapping triplet matrices and conduct the linear transformation, where $h$ is the number of heads of the feature map, and then concatenate the 12 triplet matrices after calculating the self-attention score, respectively. Finally, the high-level features $\mathbf{E}$ are obtained.}
\label{pic56}
\end{figure}

\begin{figure}[!htb]
\centering
\includegraphics[width=0.9\columnwidth]{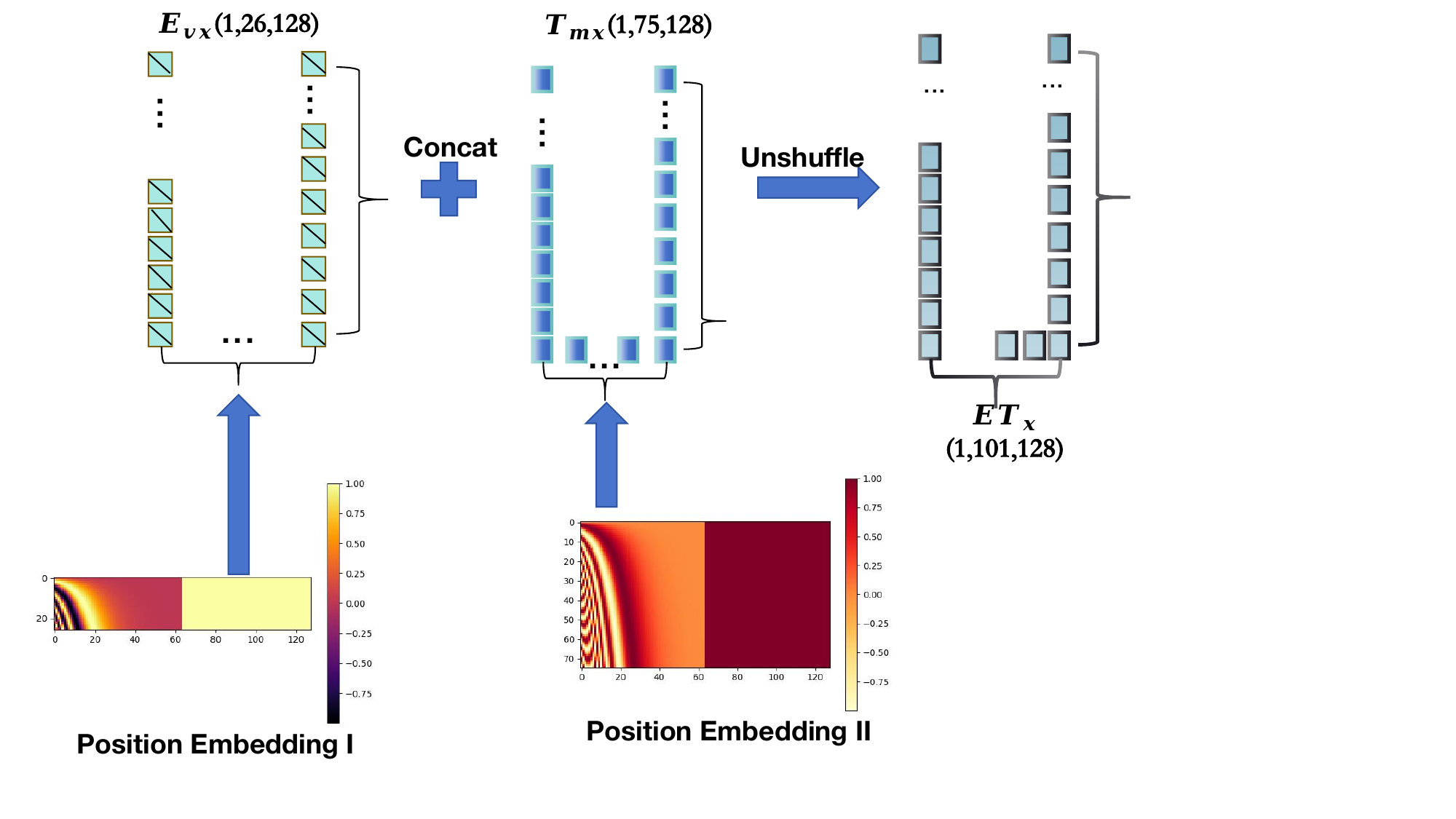} 
\caption{\textbf{The mask token operation} is to combine the token part $\mathbf{T}_{mx}$ that was previously hidden by the mask geneator with the high-level feature $\mathbf{E}_{vx}$ without disturbing the order, carry out a position embedding on each of them before the concatenation and fusion, and finally get $\mathbf{ET}_{x}$.}
\label{pic7}
\end{figure}

{\bf ViT Decoder} \quad Before the decoding operation, shown in Figure \ref{pic7}, we must combine the high-level feature information $\mathbf{E}_{vx}$ and the masked tokens $\mathbf{T}_{mx}$, and both of them should independently add the new position embedding information to form $\mathbf{ET}_{x}$. Second, we input the into the decoder $\mathbf{\psi}$ to decode the underlying features layer by layer. In each decoder block, the model first captures the global dependency through the self-attention layers, then enhances the feature expression through the feedforward network layer, and finally recovers the original input $\mathbf{XR}$ details through the upsampling layer. As shown in figure \ref{pic1}, except for the difference in the number of ViT blocks, the structure of the ViT decoder is basically the same as that of the encoder.

\section{Experiments}

\subsection{Datasets and Experimental Settings}
\textbf{Proposed Benchmark: tRGG Dataset} \quad Since there is no uniform public dataset consisting of pathological and health subjects, we proposed a new dataset, \textbf{tRGG}, by merging GaitRec~\cite{2020GaiTRec} and Gutenburg~\cite{horst2021gutenberg} to alleviate the problem of category imbalance among GaitRec alone. Concretely, GaitRec is a comprehensive large-scale dataset containing bi-lateral GRF walking trials of 2,084 patients with various musculoskeletal impairments and data from 211 healthy controls for classifying healthy/pathological gait. Nontheless, the GaitRec contains an excessive number of pathological samples (positive samples), but in practice should include a large number of healthy individuals (negative samples) and a small number of pathological subjects. Therefore, the model trained like this will not be consistent with real clinical application. To solve this problem, we introduced {\bf G}ur{\bf t}enberg which only contains GRF gait samples from 350 healthy people, and combined it with {\bf G}ait{\bf R}ec to form a new dataset, \textbf{tRGG}, to increase the number of negative samples and balance the proportion of experimental samples. In order to fully prove the self-supervised learning ability of the model in the comparative experiment, we only sampled $30\%$, $20\%$, $10\%$, $5\%$, and $1\%$ of labeled pathological samples for the contrastive experiments. Both ablation and parameter sensitivity analysis are performed at $1\%$, which makes the task more challenging. Table~\ref{tRGG} proportionally summarizes the statistics of \textbf{tRGG} datasets.

\begin{table}[htp]
\caption{\emph{Statistics of the proposed benchmark \textbf{tRGG}}. {\bf P} represents pathological samples, and {\bf H} represents healthy samples. {\bf T/V} refers to the proportion of training and validation samples \textbf{in the fine-tune stage}, and for \textbf{pre-training stage}, we dump all samples of \textbf{tRGG} into the MA\textsuperscript{2} and train them without labels.}
\smallskip
\centering
\resizebox{160pt}{!}{
\smallskip\begin{tabular}{c|c|c|c}
\hline 
\hline 
Proportions & Categories(T/V-P/H) & Count & Total  \\
\hline 
\multirow{2}*{{\bf $30\%$}}& T-P/H & 20393/16574 & 36967 \\
~ & V-P/H & 47584/16574 & 64158 \\
\hline
\multirow{2}*{{\bf $20\%$}} & T-P/H & 13595/16574 & 30169 \\
~ & V-P/H & 54382/16574 & 70956 \\
\hline
\multirow{2}*{{\bf $10\%$}} & T-P/H & 6797/16574 & 23371 \\
~ & V-P/H & 61180/16574 & 77754 \\
\hline 
\multirow{2}*{{\bf $5\%$}} & T-P/H & 3398/16574  & 19972 \\
~ & V-P/H & 64579/16754 & 81153 \\
\hline
\multirow{2}*{{\bf $1\%$}} & T-P/H & 679/16574 & 17253 \\
~ & V-P/H & 67298/16574 & 83872 \\
\hline
\end{tabular}
}
\label{tRGG}
\end{table}

\noindent \textbf{Baseline models} \quad Due to the lack of sufficient baselines to compare on the gait-based ADD task, we've added three new methods to train GRF-based deep models as new baselines (marked with * below). Totally, we compared the proposed model with six baselines of the ADD task to verify its effectiveness: 1DCNN~\cite{2021Design}, LSTM~\cite{vidya2022parkinson}, 1DCNN-LSTM~\cite{pandey2022gaitrec}, *1D-VGG19~\cite{mohan2022automatic}, *1D-Inception-ResNet~\cite{szegedy2017inception}, *ViT~\cite{alexey2020image}.

\noindent \textbf{Implementation Details} \quad  We employ the AdamW optimizer~\cite{llugsi2021comparison} using the learning rate of 0.00001 and the LR scheduler set 3 as step size, $\gamma=0.1$, and weight decay of 0.05. All experiments were conducted with two Nvidia RTX A40 GPUs, and all parameters were tuned based on the validation set. The more detailed implementation details for baselines and MA\textsuperscript{2} can be found in the supplementary materials.

\noindent \textbf{Evaluation Metrics} \quad The accuracy of binary classification was adopted to evaluate our models. First, a positive sample is a pathological sample (0), and a negative sample is a healthy sample (1). When the probability value (sigmoid result) is greater than 0.5, the prediction is a negative sample, otherwise, it is a positive sample. In more detail, the accuracy formula can be expressed as:
\begin{equation}\label{acc_cri}
acc = (TP + TN) / (TP + TN + FP + FN),
\end{equation}where true positive ($TP$) indicates the number of samples predicted by the model and actually turned positive; true negative ($TN$) indicates the number of samples predicted by the model and actually turned negative; false positive ($FP$) indicates the number of samples predicted by the model to be positive but actually negative, and false negative ($FN$) indicates the number of samples predicted by the model to be negative but actually positive. 

\begin{table}[htp]
\caption{\emph{Quantitative Comparison Experiments}. The five proportions of labeled pathological samples defined between 6 methods and our proposed MA\textsuperscript{2} on the new benchmark \textbf{tRGG}. All methods are evaluated for accuracy.}
\smallskip
\centering
\resizebox{180pt}{!}{
\smallskip\begin{tabular}{|c|c|c|c|c|c|c|}
\hline 
       & & \multicolumn{5}{|c|}{{\bf Proportions}} \\
\hline 
{\bf Datasets}& {\bf Methods}& {\bf $30\%$} & {\bf $20\%$} & {\bf $10\%$} & {\bf $5\%$} & {\bf $1\%$} \\
\hline 
\multirow{7}*{\textbf{tRGG}} & 1DCNN & 58.75 & 54.13 & 47.11 & 42.33 & 41.10 \\
\cline{2-7}
~ & LSTM & 45.83 & 41.25 & 40.10 & 41.96 & 40.20  \\
\cline{2-7}
~ & 1DCNN-LSTM & 59.90 & 58.75 & 54.17 & 44.98 & 42.73  \\
\cline{2-7}
~ & 1D-VGG19 & 59.62 & 59.58 & 59.40 & 57.87 & 55.97  \\
\cline{2-7}
~ & 1D-Inception-ResNet & 75.75 & 75.18 & 74.18 & 73.97 & 73.15  \\
\cline{2-7}
~ & ViT & 81.76 & 81.49 & 81.03 & 80.51 & 74.43  \\
\cline{2-7}
~ & MA\textsuperscript{2} & {\bf 97.87} & {\bf 96.73} & {\bf 96.26} & {\bf 94.32} & {\bf 90.91}  \\
\hline
\end{tabular}
}
\label{contrastiveI}
\end{table}

\begin{table}[htp]
\caption{{\emph{Encoder-Decoder Ablation Experiments} with number of blocks and FC layers.}}
\noindent
\begin{tabular}{c c}
\hline
Enc blocks & Acc \\
\hline
1 & 91.30  \\
2 & 91.20 \\
4 & {\bf 91.33} \\
8 & 91.02 \\
12 & 90.96 \\
\end{tabular}
\quad 
\quad 
\begin{tabular}{c c}
\hline
Dec blocks & Acc \\
\hline
1 & 91.75 \\
2 & 91.18 \\
4 & 91.70 \\
8 & 91.82 \\
12 & {\bf 92.25} \\
\end{tabular}
\quad 
\quad
\begin{tabular}{c c}
\hline
Enc fc-widths & Acc \\
\hline
128 & {\bf 92.27} \\
256 & 91.60 \\
384 & 91.10 \\
512 & 91.13 \\
768 & 90.63 \\
1024 & 90.37 \\
\end{tabular}
\quad
\quad
\begin{tabular}{c c}
\hline
Dec fc-widths & Acc \\
\hline
128 & 90.50 \\
256 & 91.35 \\
384 & 91.62 \\
512 & {\bf 91.79} \\
768 & 91.28 \\
1024 & 90.95 \\
\end{tabular}
\quad
\quad
\label{ablation_study}
\end{table}  

\subsection{Contrastive Experiments on tRGG}

\noindent \textbf{Quantitative Comparison} \quad Table~\ref{contrastiveI} summarized the overall performance of gait-based ADD methods, demonstrating that our proposed MA\textsuperscript{2} surpasses all baselines on all proportions validations. Quantitatively, MA\textsuperscript{2} achieves at least a $15\%$ improvement in accuracy with the best-performing baseline. This proves that the pre-training task of manual representation motion augmenting based on token mask stimulates the representation extraction ability of the model to some extent. That is to say, it can learn the most essential features contained in the samples even with only a few label samples. Additionally, MA\textsuperscript{2} outperforms three other powerful deep learning methods such as 1D-VGG19, 1D-Inception-ResNet, and ViT, ranging from $16.11\%$ to $20.08\%$, which is mainly due to the fact that the aforementioned baselines have not undergone a pre-training process to improve implicit representation. Even though ViT has similar global-local feature extraction capabilities and self-attention mechanisms to MA\textsuperscript{2}, MA\textsuperscript{2} can still excel with mask generator and stacked multi-layer VIT encoder-decoders. We can observe that the deterioration of 1DCNN, LSTM, and 1DCNN-LSTM with the reduction of pathological samples with labels was more stable than that of the other three baselines, indicating that layer depth and width stability are not necessarily the best.

\begin{figure}[!htb]
\centering
\includegraphics[width=0.9\columnwidth]{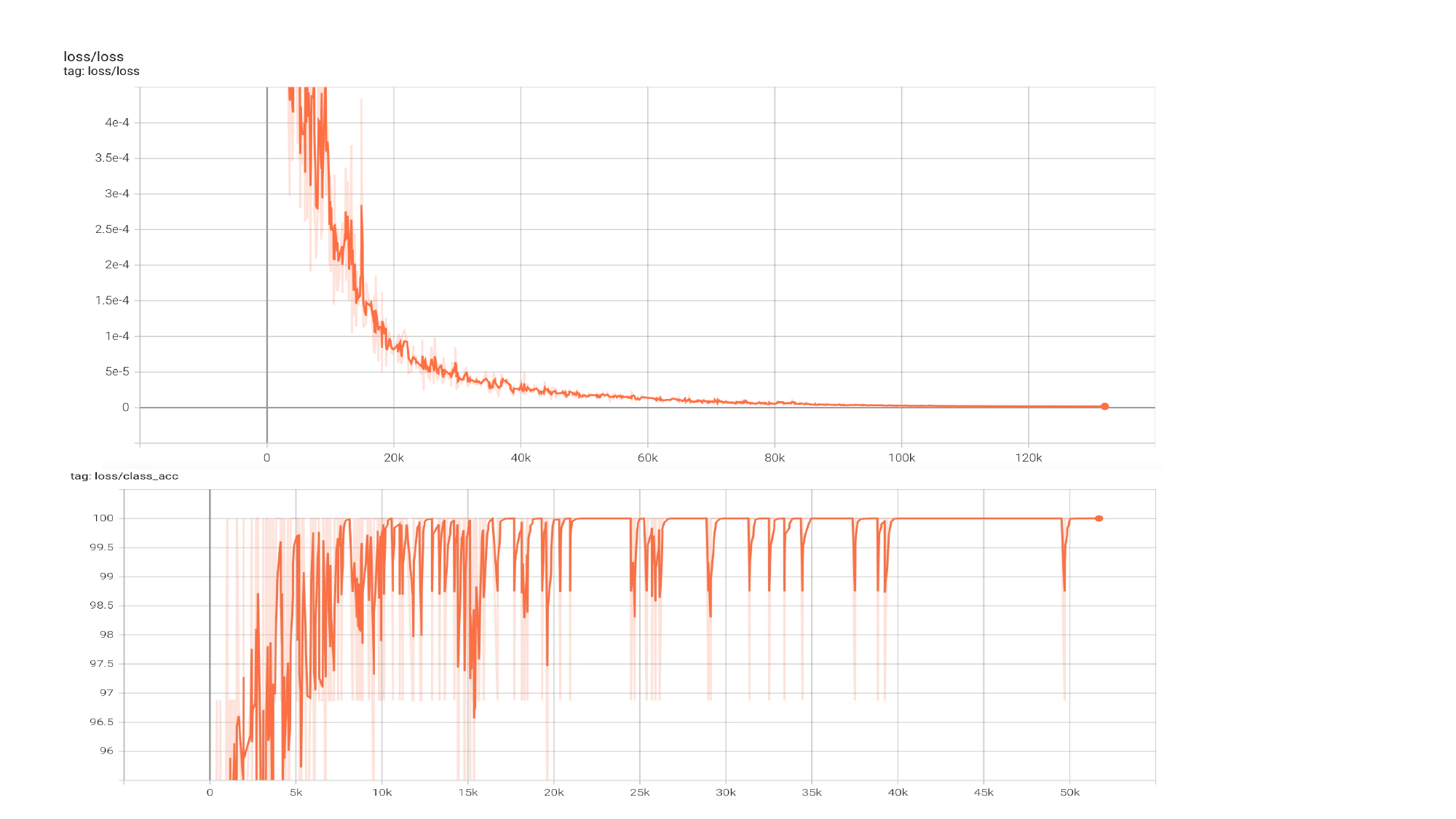} 
\caption{\textbf{Upper:}loss curve during MA\textsuperscript{2} pre-training. \textbf{Bottom:} accuracy curve during MA\textsuperscript{2} fine-tuning.}
\label{lossacc}
\end{figure}

\begin{figure}[!htb]
\centering
{\includegraphics[width=1.0\columnwidth]{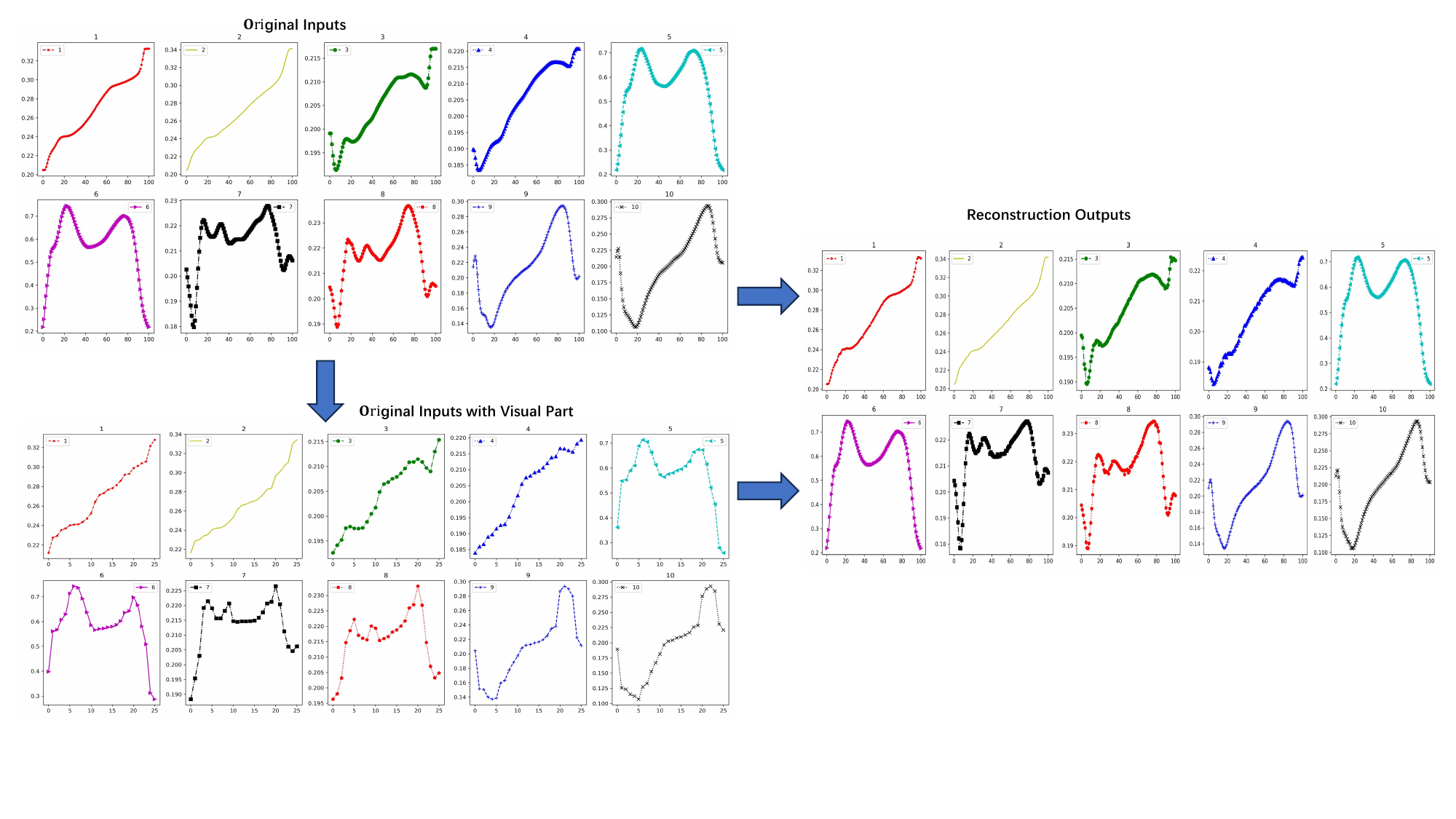}} 
\caption{\textbf{Visual reconstruction effect}. We used a mask rate of $75\%$ for this reconstruction. ``Original Inputs with Visual Part'' means that we mask out some sequence points and only show the visual part in the diagram.}
\label{vis_recs}
\end{figure}

\noindent \textbf{Qualitative Analysis} \quad As shown in the upper part of Figure~\ref{lossacc}, on 0-100k iterations, the loss in the  MA\textsuperscript{2} pre-training stage decreases relatively smoothly, even if there is a small amount of fluctuation in each interval segment. The overall loss decreases to close to 0 after 100k iterations, which is also reasonable. On the other hand, for accuracy in the fine-tuning stage in the bottom half of the picture, we chose the situation when the proportion of pathological samples was $30\%$. After 15k iterations, the overall accuracy rate of the model is roughly more than $95\%$, but there are still some fluctuations, which may indicate that some samples under the batches are difficult to classify, and we need to further optimize the algorithm later to improve their classification accuracy. Furthermore, in order to verify the reconstruction capability of our proposed MA\textsuperscript{2}, we generated the reconstruction rpocesss as shown in Figure~\ref{vis_recs}. Simply speaking, it is only one step from the original inputs to the reconstructed outputs, and the more similar the two are, the better the effect.

\subsection{Ablation Studies}

\textbf{How does eliminating the pre-training process affect performance?} \quad Since pre-training plays a key role in performance, ablation can be further assured of its effectiveness. After ablating, we found that the accuracy decreased from more than $90\%$ to $82.41\%$, so the effect can be demonstrated.

\textbf{How much does the number of encoder blocks and the width of the final linear layer affect model performance?} \quad As can be seen from the left half of the Table~\ref{ablation_study}, the impact is the biggest when the encoder block is 4, and the performance is highest when the linear layer dimension is 128.

\textbf{How much does the number of decoder blocks and the width of the final linear layer affect model performance?} \quad From the right half of Table~\ref{ablation_study}, we know that the decoder will be discarded in the validation process, so its advantages are mainly reflected in reconstruction. We only obtained through verification that the deeper the number of layers, as the impact on its performance.

\subsection{Scalable Experiments on NIH}To verify the generalization ability of the model, we introduced scalable gait datasets about GRF-based Parkinson's disease classification provided by the National Institutes of Health (NIH). This database contains measures of gait from 93 patients with idiopathic PD (mean age: 66.3 years; $63\%$ men) and 73 healthy controls (mean age: 66.3 years; $55\%$ men). The database includes the vertical ground reaction force records of subjects as they walked at their usual, self-selected pace for approximately 2 minutes on level ground. Underneath each foot were 8 sensors that measure force (in Newtons) as a function of time. The output of each of these 16 sensors has been digitized and recorded at 100 samples per second, and the records also include two signals that reflect the sum of the 8 sensor outputs for each foot. Therefore, the size of a single sample we input into the model is $X$(1,1000,18), where 1000 is the sequence length and 18 is the number of dimensions of features. We only use $10\%$ of pathological samples with labels, and 37 batches of mixed control group (healthy samples) are used for training, and the rest for testing. The experimental results are shown in Table~\ref{nih_exp}. Overall, MA\textsuperscript{2} has achieved the best performance at NIH mainly due to its asymmetric encoder-decoder structure design, and the mask operation can force the model to learn more essential, more general, and enhanced motion features, so it can show better transferable performance than other methods. We also observed that the performance of 1D-Inception-ResNet was much worse than it had been on tRGG before, possibly because the network could be deeper and wider to improve accuracy but may be more suitable for task-specific scenarios.

\begin{table}[htp]
\caption{\emph{Scalable Experiments.} Combining with other six methods. The experiment is conducted on the NIH dataset. }
\smallskip
\centering
\resizebox{130pt}{!}{
\smallskip\begin{tabular}{|c|c|c|}
\hline 
{\bf Datasets}& {\bf Methods}& {\bf Acc}  \\
\hline 
\multirow{7}*{NIH} & 1DCNN & 53.12\\
\cline{2-3}
~ & LSTM & 41.25 \\
\cline{2-3}
~ & 1DCNN-LSTM & 57.92\\
\cline{2-3}
~ & 1D-VGG19 & 58.75\\
\cline{2-3}
~ & 1D-Inception-ResNet & 38.67 \\
\cline{2-3}
~ & ViT & 61.50 \\
\cline{2-3}
~ & MA\textsuperscript{2} & {\bf 78.57}\\
\hline
\end{tabular}
}
\label{nih_exp}
\end{table}

\section{Conclusion}\label{sec5}
In this paper, we propose a self-supervised deep framework named motion augmenting autoencoder (MA\textsuperscript{2}) for automatic disease detection (ADD) based on the ground reaction force (GRF) of gait, which excels in new comprehensive benchmark datasets and handles large variations in a brand new GRF dataset, respectively. Specifically, the key component of MA\textsuperscript{2} in the pretraining process, the mask generator, can randomly mask out the sequence of tokens to augment the potential of motion representation extraction of the autoencoder. Additionally, the backbones of encoder-decoder are multi-head self-attention mechanisms (MHSAM), which allow the model to focus on information in different positions at the same time while processing the input sequence, capturing the enhanced motion information in the gait sequence more comprehensively. Meanwhile, both the mask generator and MHSAM can play a role in accelerating learning,  respectively due to masking out the vast majority of tokens and multi-head parallel computing strategies. Extensive experiments demonstrate that MA\textsuperscript{2} acquires state-of-the-art performance compared with multiple advanced deep learning methods on ADD. Furthermore, MA\textsuperscript{2} highlights its promising cross-domain capability in the scalable experiments.


\bibliographystyle{ieeetr}
\bibliography{ref}

\begin{thebibliography}{10}

\bibitem{2022Explaining}
D.~Slijepcevic, F.~Horst, S.~Lapuschkin, B.~Horsak, N.~M. Raberger, A.~Kranzl,
  W.~Samek, C.~Breiteneder, W.~I. Schollhorn, and M.~Zeppelzauer, ``Explaining
  machine learning models for clinical gait analysis,'' {\em ACM Transactions
  on Computing for Healthcare}, 2022.

\bibitem{teixeira2022automatic}
A.~L.~F. Teixeira, ``Automatic diagnosis of gait pathologies using gaitrec
  dataset,'' Master's thesis, 2022.

\bibitem{2020GaiTRec}
B.~Horsak, D.~Slijepcevic, A.~M. Raberger, C.~Schwab, and M.~Zeppelzauer,
  ``Gaitrec, a large-scale ground reaction force dataset of healthy and
  impaired gait,'' {\em Scientific Data}, vol.~7, no.~1, p.~143, 2020.

\bibitem{siddiqui2024footwear}
H.~U.~R. Siddiqui, S.~Nawaz, M.~N. Saeed, A.~A. Saleem, M.~A. Raza, A.~Raza,
  M.~A. Aslam, and S.~Dudley, ``Footwear-integrated force sensing resistor
  sensors: A machine learning approach for categorizing lower limb disorders,''
  {\em Engineering Applications of Artificial Intelligence}, vol.~127,
  p.~107205, 2024.

\bibitem{yuan2016feature}
Z.-W. Yuan and J.~Zhang, ``Feature extraction and image retrieval based on
  alexnet,'' in {\em Eighth International Conference on Digital Image
  Processing (ICDIP 2016)}, vol.~10033, pp.~65--69, SPIE, 2016.

\bibitem{horst2023modeling}
F.~Horst, D.~Slijepcevic, M.~Simak, B.~Horsak, W.~I. Sch{\"o}llhorn, and
  M.~Zeppelzauer, ``Modeling biological individuality using machine learning: A
  study on human gait,'' {\em Computational and Structural Biotechnology
  Journal}, vol.~21, pp.~3414--3423, 2023.

\bibitem{2021Design}
S.~A. Boompelli and S.~Bhattacharya, ``Design of a telemetric gait analysis
  insole and 1-d convolutional neural network to track postoperative fracture
  rehabilitation,'' in {\em 2021 IEEE 3rd Global Conference on Life Sciences
  and Technologies (LifeTech)}, 2021.

\bibitem{pandey2022gaitrec}
C.~Pandey, D.~S. Roy, R.~C. Poonia, A.~Altameem, S.~R. Nayak, A.~Verma, and
  A.~K.~J. Saudagar, ``Gaitrec-net: a deep neural network for gait disorder
  detection using ground reaction force,'' {\em PPAR research}, vol.~2022,
  no.~1, p.~9355015, 2022.

\bibitem{yun2023exploratory}
R.~Yun, M.~Salama, and L.~Elrefaei, ``An exploratory study on the effect of
  applying various artificial neural networks to the classification of lower
  limb injury,'' {\em Turkish Journal of Electrical Engineering and Computer
  Sciences}, vol.~31, no.~2, pp.~448--461, 2023.

\bibitem{yoon2019time}
J.~Yoon, D.~Jarrett, and M.~Van~der Schaar, ``Time-series generative
  adversarial networks,'' {\em Advances in Neural Information Processing
  Systems}, vol.~32, 2019.

\bibitem{eltoukhy2017prediction}
M.~Eltoukhy, C.~Kuenze, M.~S. Andersen, J.~Oh, and J.~Signorile, ``Prediction
  of ground reaction forces for parkinson's disease patients using a
  kinect-driven musculoskeletal gait analysis model,'' {\em Medical Engineering
  \& Physics}, vol.~50, pp.~75--82, 2017.

\bibitem{al2023pad}
A.~Al-Ramini, ``Pad diagnosis and estimation of treatment effectiveness using
  machine learning,'' {\em Diss. Dr. Doc. Univ. Neb.-Linc. 2024}, 2023.

\bibitem{sreeraman2023drug}
S.~Sreeraman, M.~P. Kannan, R.~B. Singh~Kushwah, V.~Sundaram, A.~Veluchamy,
  A.~Thirunavukarasou, and K.~M. Saravanan, ``Drug design and disease
  diagnosis: the potential of deep learning models in biology,'' {\em Current
  Bioinformatics}, vol.~18, no.~3, pp.~208--220, 2023.

\bibitem{horst2021gutenberg}
F.~Horst, D.~Slijepcevic, M.~Simak, and W.~I. Sch{\"o}llhorn, ``Gutenberg gait
  database, a ground reaction force database of level overground walking in
  healthy individuals,'' {\em Scientific Data}, vol.~8, no.~1, p.~232, 2021.

\bibitem{jlassi2024effect}
O.~Jlassi and P.~C. Dixon, ``The effect of time normalization and biomechanical
  signal processing techniques of ground reaction force curves on deep-learning
  model performance,'' {\em Journal of Biomechanics}, vol.~168, p.~112116,
  2024.

\bibitem{andrei2019parkinson}
A.-G. Andrei, A.-M. T{\u{a}}uțan, and B.~Ionescu, ``Parkinson’s disease
  detection from gait patterns,'' in {\em 2019 E-Health and Bioengineering
  Conference (EHB)}, pp.~1--4, IEEE, 2019.

\bibitem{fuadah2022parkinson}
Y.~N. Fuadah, F.~F. Taliningsih, I.~Wijayanto, N.~K.~C. Pratiwi, and S.~Rizal,
  ``Parkinson’s disease detection based on gait analysis of vertical ground
  reaction force using signal processing with machine learning,'' in {\em
  Proceedings of the 2nd International Conference on Electronics, Biomedical
  Engineering, and Health Informatics: ICEBEHI 2021, 3--4 November, Surabaya,
  Indonesia}, pp.~253--264, Springer, 2022.

\bibitem{alharthi2020gait}
A.~S. Alharthi, A.~J. Casson, and K.~B. Ozanyan, ``Gait spatiotemporal signal
  analysis for parkinson’s disease detection and severity rating,'' {\em IEEE
  Sensors Journal}, vol.~21, no.~2, pp.~1838--1848, 2020.

\bibitem{chakraborty2022musculoskeletal}
J.~Chakraborty, S.~Upadhyay, and A.~Nandy, ``Musculoskeletal injury recovery
  assessment using gait analysis with ground reaction force sensor,'' {\em
  Medical Engineering \& Physics}, vol.~103, p.~103788, 2022.

\bibitem{harithasan2023review}
D.~Harithasan and N.~A.~B. Abd~Razak, ``A review of the analysis of ground
  reaction force among adults with lower limb problems,'' {\em Journal Sains
  Kesihatan Malaysia}, vol.~21, no.~2, pp.~1--10, 2023.

\bibitem{wu2020imu}
C.-C. Wu, Y.-T. Wen, and Y.-J. Lee, ``Imu sensors beneath walking surface for
  ground reaction force prediction in gait,'' {\em IEEE Sensors Journal},
  vol.~20, no.~16, pp.~9372--9376, 2020.

\bibitem{sivakumar2022instrumented}
A.~Sivakumar, K.~Bennett, M.~Rickman, and D.~Thewlis, ``An instrumented walker
  in three-dimensional gait analysis: Improving musculoskeletal estimates in
  the lower limb mobility impaired,'' {\em Gait \& Posture}, vol.~93,
  pp.~142--145, 2022.

\bibitem{ng2011sparse}
A.~Ng {\em et~al.}, ``Sparse autoencoder,'' {\em CS294A Lecture Notes},
  vol.~72, no.~2011, pp.~1--19, 2011.

\bibitem{cheng2018deep}
Z.~Cheng, H.~Sun, M.~Takeuchi, and J.~Katto, ``Deep convolutional
  autoencoder-based lossy image compression,'' in {\em 2018 Picture Coding
  Symposium (PCS)}, pp.~253--257, IEEE, 2018.

\bibitem{doersch2016tutorial}
C.~Doersch, ``Tutorial on variational autoencoders,'' {\em ArXiv Preprint
  ArXiv:1606.05908}, 2016.

\bibitem{vincent2010stacked}
P.~Vincent, H.~Larochelle, I.~Lajoie, Y.~Bengio, P.-A. Manzagol, and L.~Bottou,
  ``Stacked denoising autoencoders: Learning useful representations in a deep
  network with a local denoising criterion.,'' {\em Journal of Machine Learning
  Research}, vol.~11, no.~12, 2010.

\bibitem{he2022masked}
K.~He, X.~Chen, S.~Xie, Y.~Li, P.~Doll{\'a}r, and R.~Girshick, ``Masked
  autoencoders are scalable vision learners,'' in {\em Proceedings of the
  IEEE/CVF Conference on Computer Vision and Pattern Recognition},
  pp.~16000--16009, 2022.

\bibitem{ioffe2015batch}
S.~Ioffe and C.~Szegedy, ``Batch normalization: Accelerating deep network
  training by reducing internal covariate shift,'' in {\em International
  Conference on Machine Learning}, pp.~448--456, PMLR, 2015.

\bibitem{alexey2020image}
D.~Alexey, ``An image is worth 16x16 words: Transformers for image recognition
  at scale,'' {\em ArXiv Preprint ArXiv: 2010.11929}, 2020.

\bibitem{vidya2022parkinson}
B.~Vidya and P.~Sasikumar, ``Parkinson’s disease diagnosis and stage
  prediction based on gait signal analysis using emd and cnn--lstm network,''
  {\em Engineering Applications of Artificial Intelligence}, vol.~114,
  p.~105099, 2022.

\bibitem{mohan2022automatic}
R.~Mohan, S.~Kadry, V.~Rajinikanth, A.~Majumdar, and O.~Thinnukool, ``Automatic
  detection of tuberculosis using vgg19 with seagull-algorithm,'' {\em Life},
  vol.~12, no.~11, p.~1848, 2022.

\bibitem{szegedy2017inception}
C.~Szegedy, S.~Ioffe, V.~Vanhoucke, and A.~Alemi, ``Inception-v4,
  inception-resnet and the impact of residual connections on learning,'' in
  {\em Proceedings of the AAAI Conference on Artificial Intelligence}, vol.~31,
  2017.

\bibitem{llugsi2021comparison}
R.~Llugsi, S.~El~Yacoubi, A.~Fontaine, and P.~Lupera, ``Comparison between
  adam, adamax and adam w optimizers to implement a weather forecast based on
  neural networks for the andean city of quito,'' in {\em 2021 IEEE Fifth
  Ecuador Technical Chapters Meeting (ETCM)}, pp.~1--6, IEEE, 2021.

\end{thebibliography}

\end{document}